\documentclass[twocolumn,prb,amsmath,amssymb,floatfix,superscriptaddress,showpacs]{revtex4}
\usepackage{graphicx}
\usepackage{dcolumn}
\usepackage{bm}

\usepackage{times}
\begin{document}


\title{ Response of polar nanoregions in $68\%$Pb(Mg$_{1/3}$Nb$_{2/3}$)O$_3$-$32\%$PbTiO$_3$ to a [001] electric field}
\author{Jinsheng Wen\footnote{Author to whom correspondence should be addressed; electronic mail:jwen@bnl.gov.}}
\affiliation{Condensed Matter Physics and Materials Science
Department, Brookhaven National Laboratory, Upton, New York 11973,
USA} \affiliation{Department of Materials Science and Engineering,
Stony Brook University, Stony Brook, New York 11794, USA}
\author{Guangyong Xu}
\affiliation{Condensed Matter Physics and Materials Science
Department, Brookhaven National Laboratory, Upton, New York 11973,
USA}
\author{C. Stock}
\affiliation{ISIS Facility, Rutherford Appleton Laboratory, Didcot,
OX11 0QX, UK}
\author{P. M. Gehring}
\affiliation{NIST Center for Neutron Research, National Institute of
Standards and Technology, Gaithersburg, Maryland 20899, USA}
\date{\today}

\begin{abstract}
We report neutron diffuse scattering measurements on a single
crystal of $68\%$Pb(Mg$_{1/3}$Nb$_{2/3}$)O$_3$-$32\%$PbTiO$_3$.
Strong diffuse scattering is observed at low temperatures. An
external field applied along the [001] direction affects the diffuse
scattering in the (HK0) plane significantly, suggesting a
redistribution occurs between polar nanoregions of different
polarizations perpendicular to the field. By contrast, the [001]
field has no effect on the diffuse scattering in the (HOL) and (0KL)
zones.

\end{abstract}


\maketitle

The complex perovskites
$(1-x)$Pb(Mg$_{1/3}$Nb$_{2/3}$)O$_3$-$x$PbTiO$_3$(PMN-$x$PT) and
$(1-x)$Pb(Zn$_{1/3}$Nb$_{2/3}$)O$_3$-$x$PbTiO$_3$(PZN-$x$PT) are of
great interest because of their promising piezoelectric
properties.~\cite{PZT1,Uchino,Service} They are prototypical
ferroelectric relaxors (relaxors hereafter) that have large and
strongly frequency-dependent dielectric constants, which peak
broadly in temperature.~\cite{GY1,PZT1} A unique property of
relaxors is the existence of polar nanoregions (PNR), a concept
first proposed by Burns and Dacol in 1983.~\cite{Burns} The PNR
start to form on cooling at the "Burns temperature" T$_d$, continue
to grow with decreasing temperature, and can be directly probed by
neutron/x-ray.~\cite{W,PZN_diffuse,PZN_diffuse2,PZN_diffuse3,Hiro_diffuse,PMN_diffuse2,PMN_xraydiffuse,PMN_xraydiffuse2}

Recent work on PZN-4.5\%PT suggests a close connection between PNR
and the ultra-high piezoelectric response in
relaxors.~\cite{Xu_PZNNATUREMAT} Previous studies have confirmed
that diffuse scattering from PNR disappears for compositions on the
ferroelectric side of the phase
diagram~\cite{matsuura:144107,PMN-60PT}, while the integrated
diffuse scattering intensity appears to reach a maximum near the
morphotropic phase boundary (MPB)~\cite{matsuura:144107}. The
application of an external field along [111] redistributes the PNR,
resulting in a change in the diffuse scattering
patterns;~\cite{Xu_nm1,xu:104110,Xu_PZNNATUREMAT} however the effect
of external field along [001] is not yet fully
understood~\cite{PZN-8PT}. To understand the role PNR play in the
piezoelectric response of relaxor systems, we believe it is now
becoming more important to understand how they behave within the MPB
region where the piezoelectric properties are optimal in both
PMN-$x$PT and PZN-$x$PT~\cite{ferrobook,PZN_phase2,shrout1}, and in
particular how they respond to an external field along [001], which
is the poling direction that produces the greatest piezoelectric
effect.

In this Letter, we present neutron scattering results on a
PMN-32\%PT single crystal, a composition that lies inside the MPB.
Our results show that (i) there is strong diffuse scattering in this
compound with a spatial distribution similar to that observed in
other PMN-$x$PT and PZN-$x$PT
systems~\cite{GXU3D,PMN_diffuse,PZN_diffuse3,PMN_xraydiffuse,PMN_xraydiffuse2,Hiro_diffuse};
(ii) the diffuse scattering in the (HK0) plane responds to a
moderate external field (E = 2~kV/cm) applied along [001], which is
probably associated with low symmetry local structures induced by
the field and internal strain in the compound; (iii) the field does
not affect the diffuse scattering in the (H0L) and (0KL) zones.

The crystal has a rectangular shape, dimensions of
$10\times10\times2$~mm$^3$ with six \{100\} surfaces, and was
provided by TRS Ceramic. Neutron scattering experiments were carried
out on the triple-axis spectrometer BT9 located at the NIST Center
for Neutron Research (NCNR) using beam collimations of
40'-40'-S-40'-80' (S=sample) with fixed initial and final neutron
energy of 14.7~meV. An electric field of 2~kV/cm was applied along
[001] during field-cooled (FC) measurements.

\begin{figure}[ht]
\includegraphics[width=\linewidth]{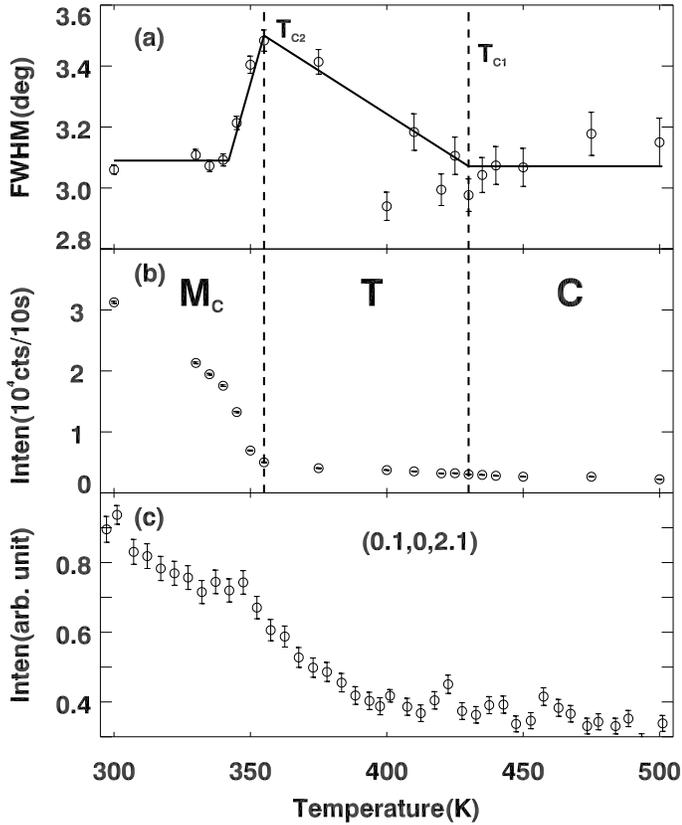}
\caption{Temperature dependence of the (a) (300) peak longitudinal
FWHM, (b) (300) peak intensity, and (c) diffuse scattering intensity
at (0.1,0,2.1). Solid lines are guides to the eye. Dashed lines
indicate the phase transition temperatures. Error bars in (a) are
obtained by least-square fitting the data with Gaussian functions,
and those in (b) and (c) represent the square root of the counts.
The bump in (c) at $\sim$355~K is due to critical scattering.}
\label{fig:1}
\end{figure}

In Fig.~\ref{fig:1}(a) and (b), the (300) Bragg peak longitudinal
full width at half maximum (FWHM) and intensity are plotted. Two
phase transitions occur at T$_{C1} \sim 430$~K (cubic (C) to
tetragonal (T)) and T$_{C2} \sim 355$~K (T to monoclinic (M)). These
results are consistent with previous
results~\cite{PMN_phase,hucao1,gvasaliya}, and confirm that the
composition of the sample lies inside the MPB.

\begin{figure}[ht]
\includegraphics[width=\linewidth]{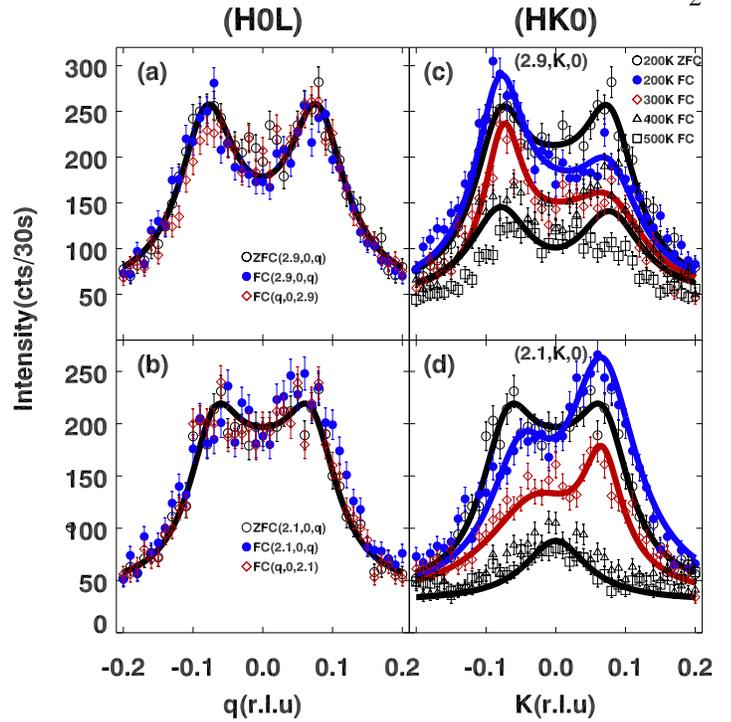}
\caption{(Color online) Linear scans of the diffuse scattering
intensity measured in the (H0L) plane at 200~K around (a) (300)
(ZFC, open circles; FC closed circles) and (003) (FC diamonds); (b)
(200) (ZFC, open circles; FC, filled circles) and (002) (FC,
diamonds). (c) and (d) are linear scans of the diffuse scattering
along (2.9, K, 0) and (2.1, K, 0) at different temperatures in the
(HK0) plane. Lines are guides to the eye. Error bars represent the
square root of the counts.} \label{fig:2}
\end{figure}

In our PMN-32\%PT sample, we observed strong diffuse scattering,
which increases monotonically with cooling as demonstrated in
Fig.~\ref{fig:1}(c). In Fig.~\ref{fig:2}(a) and (b), we show
selected linear scans offset from the (300), (003), (200) and (002)
Bragg peaks in the (H0L) plane under zero-field-cooled (ZFC) and FC
conditions at 200~K. When we compare ZFC and FC results, it is clear
that in the (H0L) zone, a [001] field has no detectable effect on
the diffuse scattering. Based on earlier models~\cite{GXU3D}, we
suggest that in this compound PNR with polarizations not
perpendicular to the [001] field (in this case, [101], [011],
[$\bar{1}01$], [$0\bar{1}1$]) are not affected.

Fig.~\ref{fig:2}(c) and (d) show similar linear scans of the diffuse
scattering measured near (200) and (300) at different temperatures
in the (HK0) zone, where the [001] electric field is now
perpendicular to the scattering plane. Here, the FC and ZFC data now
show clear differences. When ZFC at 200~K, the diffuse scattering
exhibits a symmetric double-peaked profile for both (2.9,K,0) and
(2.1,K,0) scans. When FC at T $\agt$ 400~K, where the diffuse
scattering is still weak, the field effect is not apparent and the
diffuse scattering remains symmetric in shape; at lower temperatures
the diffuse scattering becomes more intense, and the diffuse profile
becomes asymmetric.

\begin{figure}[ht]
\includegraphics[width=\linewidth,height=1.1\linewidth]{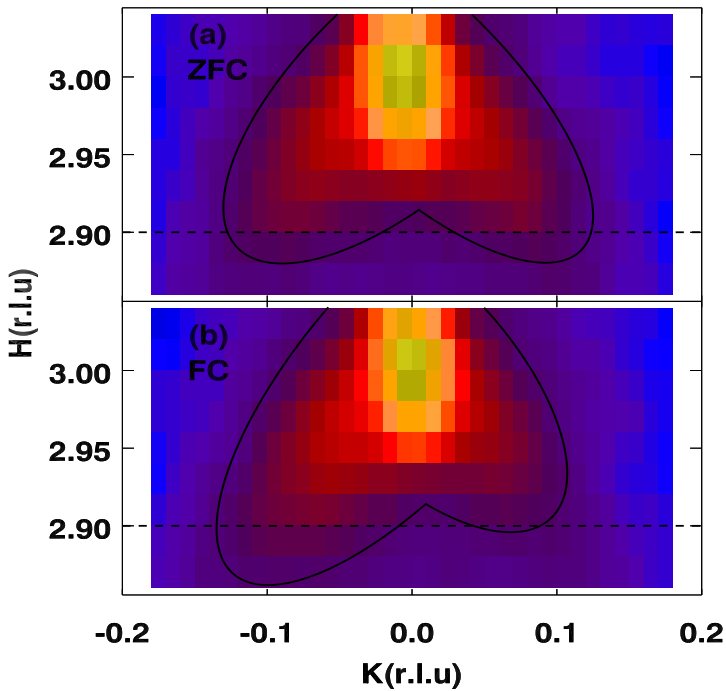}
\caption{(Color online) Contour maps of diffuse scattering around
(300) at 200 K, (a) ZFC, (b) FC. Solid lines are guides to the eye.
Dashed lines indicate the scans in Fig.~\ref{fig:2}(c) at 200~K.}
\label{fig:3}
\end{figure}

These features can be more clearly seen in the diffuse scattering
contour maps in Fig.~\ref{fig:3}. Fig.~\ref{fig:3}(a) shows that at
200~K, under ZFC conditions, there is strong diffuse scattering
having a symmetric butterfly shape with one wing  along [110] and
the other along [$\bar1$10], centered at (300). We are only able to
measure the bottom part of this butterfly because of mechanical
restrictions on the Q-range. Under FC conditions, some intensity
from the wing along [$\bar1$10] is shifted to the other wing along
[110], creating the asymmetry.

We have performed multiple FC sequences and these results are
reproducible. Also note that the change in diffuse scattering in the
(HK0) plane persists even after the removal of the external field
below T$_C$. Only by heating the sample to high temperature (500~K)
and cooling in zero field can one remove the field effect. This is
similar to that measured in PZN-$x$PT
samples,~\cite{Xu_new,xu:104110} where the change of the diffuse
scattering is believed to be associated with the formation and
change of ferroelectric domains. However, there are some differences
between the effect of external field along [001] and [111] in that,
(i) a [001] field only affects some of the PNR (those with
polarizations in the (H0L) and (0KL) planes are not affected); (ii)
the magnitude of the redistribution (enhancement/suppression) is
smaller for a [001] field compared to that for a [111] field. To
understand the effect of domain change on the PNR, one needs to take
into consideration the low temperature structure of PMN-32\%PT.
Previous x-ray diffraction work suggests that the system enters a
monoclinic-C (M$_C$) phase when cooled under a [001]
field.~\cite{hucao1} Under ZFC conditions, any domain effect will be
averaged over the multi-domain state. On the other hand, when cooled
under a field along [001], the domain structure becomes more
organized, where (it is believed that) the c$^*$ is fixed along the
field direction. Our results suggest that PNR with [101], [011],
[$\bar{1}01$], and [$0\bar{1}1$] polarizations are not affected when
the phase transition into this c$^*$ fixed monoclinic domain state
occurs, whereas those PNR with [110] and [1$\bar{1}$0] polarizations
are affected.

In an M$_C$ phase, either the cubic a-axis or cubic b-axis is tilted
up/down towards the c-axis. In a perfect system, one would have
equal numbers of all four different M$_C$ domains. (See Fig.~2 in
Ref.~\onlinecite{skin1}.) In reality, this may not be the case due
to external/internal strains, growth conditions, etc. Nevertheless,
the asymmetric butterfly-shaped diffuse scattering suggests that in
this c$^*$ fixed M$_C$ phase, there are more PNR with [1$\bar{1}$0]
polarizations (which yield diffuse scattering along the [110]
direction) than those with [110] polarizations. This apparently
cannot be naively explained by any distribution of c$^*$ fixed M$_C$
domains, since [110] and [1$\bar{1}$0] directions are equivalent in
any of the four M$_C$ domains. Our speculation is that the true
(local) symmetry of these M$_C$ domains could be even lower (e.g.
triclinic), as predicted by Vanderbilt {\it et al.} using
higher-order Devonshire theory,~\cite{vanderbilt} so that [110] and
[1$\bar{1}$0] directions are different in these domains and a
preference for PNR with these polarizations within each domain is
realized, and observed when ferroelectric domain distribution
changes.

In summary, we have performed neutron diffuse scattering
measurements on a PMN-32\%PT single crystal under ZFC and FC
conditions. It is shown that strongly anisotropic diffuse scattering
exists, which increases monotonically with cooling. These results
suggest that PNR do exist and behave very similarly for this
composition inside the MPB region compared to those on the left side
(relaxor side) of the PT doping phase diagram previously studied.
The shape of the diffuse scattering can be modified by an external
field of moderate strength along [001]. The true nature of this
field effect on the PNR is not yet fully understood, but we
conjecture that it might be related to the formation/redistribution
of ferroelectric domains having a symmetry even lower than
monoclinic, in which the PNR reside.

We would like to thank Y. Chen and W. Ratcliff for stimulating
discussions. The work at Brookhaven National Laboratory was
supported by the U.S. Department of Energy under contract
No.~DE-AC02-98CH10886.


\begin{thebibliography}{31}
\expandafter\ifx\csname
natexlab\endcsname\relax\def\natexlab#1{#1}\fi
\expandafter\ifx\csname bibnamefont\endcsname\relax
  \def\bibnamefont#1{#1}\fi
\expandafter\ifx\csname bibfnamefont\endcsname\relax
  \def\bibfnamefont#1{#1}\fi
\expandafter\ifx\csname citenamefont\endcsname\relax
  \def\citenamefont#1{#1}\fi
\expandafter\ifx\csname url\endcsname\relax
  \def\url#1{\texttt{#1}}\fi
\expandafter\ifx\csname urlprefix\endcsname\relax\def\urlprefix{URL
}\fi \providecommand{\bibinfo}[2]{#2}
\providecommand{\eprint}[2][]{\url{#2}}

\bibitem[{\citenamefont{Park and Shrout}(1997)}]{PZT1}
\bibinfo{author}{\bibfnamefont{S.-E.} \bibnamefont{Park}} \bibnamefont{and}
  \bibinfo{author}{\bibfnamefont{T.~R.} \bibnamefont{Shrout}},
  \bibinfo{journal}{J. Appl. Phys.} \textbf{\bibinfo{volume}{82}},
  \bibinfo{pages}{1804} (\bibinfo{year}{1997}).

\bibitem[{\citenamefont{Uchino}(1996)}]{Uchino}
\bibinfo{author}{\bibfnamefont{K.}~\bibnamefont{Uchino}},
  \emph{\bibinfo{title}{Piezoelectric actuators and ultrasonic motors}}
  (\bibinfo{publisher}{Kluwer Academic, Boston}, \bibinfo{year}{1996}).

\bibitem[{\citenamefont{Service}(1997)}]{Service}
\bibinfo{author}{\bibfnamefont{R.~F.} \bibnamefont{Service}},
  \bibinfo{journal}{Science} \textbf{\bibinfo{volume}{275}},
  \bibinfo{pages}{1878} (\bibinfo{year}{1997}).

\bibitem[{\citenamefont{Smolensky and Agranovskaya}(1960)}]{GY1}
\bibinfo{author}{\bibfnamefont{G.}~\bibnamefont{Smolensky}} \bibnamefont{and}
  \bibinfo{author}{\bibfnamefont{A.~I.} \bibnamefont{Agranovskaya}},
  \bibinfo{journal}{Sov. Phys. Solid State} \textbf{\bibinfo{volume}{1}},
  \bibinfo{pages}{1429} (\bibinfo{year}{1960}).

\bibitem[{\citenamefont{Burns and Dacol}(1983)}]{Burns}
\bibinfo{author}{\bibfnamefont{G.}~\bibnamefont{Burns}} \bibnamefont{and}
  \bibinfo{author}{\bibfnamefont{F.~H.} \bibnamefont{Dacol}},
  \bibinfo{journal}{Phys. Rev. B} \textbf{\bibinfo{volume}{28}},
  \bibinfo{pages}{2527} (\bibinfo{year}{1983}).

\bibitem[{\citenamefont{Wakimoto et~al.}(2006)\citenamefont{Wakimoto, Samara,
  Grubbs, Venturini, Boatner, G.Xu, Shirane, and Lee}}]{W}
\bibinfo{author}{\bibfnamefont{S.}~\bibnamefont{Wakimoto}},
  \bibinfo{author}{\bibfnamefont{G.}~\bibnamefont{Samara}},
  \bibinfo{author}{\bibfnamefont{R.}~\bibnamefont{Grubbs}},
  \bibinfo{author}{\bibfnamefont{E.}~\bibnamefont{Venturini}},
  \bibinfo{author}{\bibfnamefont{L.}~\bibnamefont{Boatner}},
  \bibinfo{author}{\bibnamefont{G.Xu}},
  \bibinfo{author}{\bibfnamefont{G.}~\bibnamefont{Shirane}}, \bibnamefont{and}
  \bibinfo{author}{\bibfnamefont{S.-H.} \bibnamefont{Lee}},
  \bibinfo{journal}{Phys. Rev. B} \textbf{\bibinfo{volume}{74}},
  \bibinfo{pages}{054101} (\bibinfo{year}{2006}).

\bibitem[{\citenamefont{La-Orauttapong
  et~al.}(2001)\citenamefont{La-Orauttapong, Toulouse, Robertson, and
  Ye}}]{PZN_diffuse}
\bibinfo{author}{\bibfnamefont{D.}~\bibnamefont{La-Orauttapong}},
  \bibinfo{author}{\bibfnamefont{J.}~\bibnamefont{Toulouse}},
  \bibinfo{author}{\bibfnamefont{J.~L.} \bibnamefont{Robertson}},
  \bibnamefont{and} \bibinfo{author}{\bibfnamefont{Z.-G.} \bibnamefont{Ye}},
  \bibinfo{journal}{Phys. Rev. B} \textbf{\bibinfo{volume}{64}},
  \bibinfo{pages}{212101} (\bibinfo{year}{2001}).

\bibitem[{\citenamefont{La-Orauttapong
  et~al.}(2003)\citenamefont{La-Orauttapong, Toulouse, Ye, Chen, Erwin, and
  Robertson}}]{PZN_diffuse2}
\bibinfo{author}{\bibfnamefont{D.}~\bibnamefont{La-Orauttapong}},
  \bibinfo{author}{\bibfnamefont{J.}~\bibnamefont{Toulouse}},
  \bibinfo{author}{\bibfnamefont{Z.-G.} \bibnamefont{Ye}},
  \bibinfo{author}{\bibfnamefont{W.}~\bibnamefont{Chen}},
  \bibinfo{author}{\bibfnamefont{R.}~\bibnamefont{Erwin}}, \bibnamefont{and}
  \bibinfo{author}{\bibfnamefont{J.~L.} \bibnamefont{Robertson}},
  \bibinfo{journal}{Phys. Rev. B} \textbf{\bibinfo{volume}{67}},
  \bibinfo{pages}{134110} (\bibinfo{year}{2003}).

\bibitem[{\citenamefont{Hlinka et~al.}(2003)\citenamefont{Hlinka, Kamba,
  Petzelt, Kulda, Randall, and Zhang}}]{PZN_diffuse3}
\bibinfo{author}{\bibfnamefont{J.}~\bibnamefont{Hlinka}},
  \bibinfo{author}{\bibfnamefont{S.}~\bibnamefont{Kamba}},
  \bibinfo{author}{\bibfnamefont{J.}~\bibnamefont{Petzelt}},
  \bibinfo{author}{\bibfnamefont{J.}~\bibnamefont{Kulda}},
  \bibinfo{author}{\bibfnamefont{C.~A.} \bibnamefont{Randall}},
  \bibnamefont{and} \bibinfo{author}{\bibfnamefont{S.~J.} \bibnamefont{Zhang}},
  \bibinfo{journal}{J. Phys.: Condens. Matter} \textbf{\bibinfo{volume}{15}},
  \bibinfo{pages}{4249} (\bibinfo{year}{2003}).

\bibitem[{\citenamefont{Hiraka et~al.}(2004)\citenamefont{Hiraka, Lee, Gehring,
  Xu, and Shirane}}]{Hiro_diffuse}
\bibinfo{author}{\bibfnamefont{H.}~\bibnamefont{Hiraka}},
  \bibinfo{author}{\bibfnamefont{S.-H.} \bibnamefont{Lee}},
  \bibinfo{author}{\bibfnamefont{P.~M.} \bibnamefont{Gehring}},
  \bibinfo{author}{\bibfnamefont{G.}~\bibnamefont{Xu}}, \bibnamefont{and}
  \bibinfo{author}{\bibfnamefont{G.}~\bibnamefont{Shirane}},
  \bibinfo{journal}{Phys. Rev. B} \textbf{\bibinfo{volume}{70}},
  \bibinfo{pages}{184105} (\bibinfo{year}{2004}).

\bibitem[{\citenamefont{Dkhil et~al.}(2001)\citenamefont{Dkhil, Kiat, Calvarin,
  Baldinozzi, Vakhrushev, and Suard}}]{PMN_diffuse2}
\bibinfo{author}{\bibfnamefont{B.}~\bibnamefont{Dkhil}},
  \bibinfo{author}{\bibfnamefont{J.~M.} \bibnamefont{Kiat}},
  \bibinfo{author}{\bibfnamefont{G.}~\bibnamefont{Calvarin}},
  \bibinfo{author}{\bibfnamefont{G.}~\bibnamefont{Baldinozzi}},
  \bibinfo{author}{\bibfnamefont{S.~B.} \bibnamefont{Vakhrushev}},
  \bibnamefont{and} \bibinfo{author}{\bibfnamefont{E.}~\bibnamefont{Suard}},
  \bibinfo{journal}{Phys. Rev. B} \textbf{\bibinfo{volume}{65}},
  \bibinfo{pages}{024104} (\bibinfo{year}{2001}).

\bibitem[{\citenamefont{You and Zhang}(1997)}]{PMN_xraydiffuse}
\bibinfo{author}{\bibfnamefont{H.}~\bibnamefont{You}} \bibnamefont{and}
  \bibinfo{author}{\bibfnamefont{Q.~M.} \bibnamefont{Zhang}},
  \bibinfo{journal}{Phys. Rev. Lett.} \textbf{\bibinfo{volume}{79}},
  \bibinfo{pages}{3950} (\bibinfo{year}{1997}).

\bibitem[{\citenamefont{Takesue et~al.}(2001)\citenamefont{Takesue, Fujii, and
  You}}]{PMN_xraydiffuse2}
\bibinfo{author}{\bibfnamefont{N.}~\bibnamefont{Takesue}},
  \bibinfo{author}{\bibfnamefont{Y.}~\bibnamefont{Fujii}}, \bibnamefont{and}
  \bibinfo{author}{\bibfnamefont{H.}~\bibnamefont{You}},
  \bibinfo{journal}{Phys. Rev. B} \textbf{\bibinfo{volume}{64}},
  \bibinfo{pages}{184112} (\bibinfo{year}{2001}).

\bibitem[{\citenamefont{{Guangyong Xu} et~al.}()\citenamefont{{Guangyong Xu},
  {Jinsheng Wen}, Stock, and Gehring}}]{Xu_PZNNATUREMAT}
\bibinfo{author}{\bibnamefont{{Guangyong Xu}}},
  \bibinfo{author}{\bibnamefont{{Jinsheng Wen}}},
  \bibinfo{author}{\bibfnamefont{C.}~\bibnamefont{Stock}}, \bibnamefont{and}
  \bibinfo{author}{\bibfnamefont{P.~M.} \bibnamefont{Gehring}},
  \bibinfo{journal}{Nature Mater.} \textbf{\bibinfo{volume}{7}},
  \bibinfo{pages}{562} (\bibinfo{year}{2008}).

\bibitem[{\citenamefont{Matsuura et~al.}(2006)\citenamefont{Matsuura, Hirota,
  Gehring, Ye, Chen, and Shirane}}]{matsuura:144107}
\bibinfo{author}{\bibfnamefont{M.}~\bibnamefont{Matsuura}},
  \bibinfo{author}{\bibfnamefont{K.}~\bibnamefont{Hirota}},
  \bibinfo{author}{\bibfnamefont{P.~M.} \bibnamefont{Gehring}},
  \bibinfo{author}{\bibfnamefont{Z.-G.} \bibnamefont{Ye}},
  \bibinfo{author}{\bibfnamefont{W.}~\bibnamefont{Chen}}, \bibnamefont{and}
  \bibinfo{author}{\bibfnamefont{G.}~\bibnamefont{Shirane}},
  \bibinfo{journal}{Phys. Rev. B} \textbf{\bibinfo{volume}{74}},
  \bibinfo{pages}{144107} (\bibinfo{year}{2006}).

\bibitem[{\citenamefont{Stock et~al.}(2006)\citenamefont{Stock, Ellis,
  Swainson, Xu, Hiraka, Zhong, Luo, Zhao, Viehland, Birgeneau, and Shirane}}]{PMN-60PT}
\bibinfo{author}{\bibfnamefont{C.}~\bibnamefont{Stock}},
  \bibinfo{author}{\bibfnamefont{D.}~\bibnamefont{Ellis}},
  \bibinfo{author}{\bibfnamefont{I.~P.} \bibnamefont{Swainson}},
  \bibinfo{author}{\bibfnamefont{Guangyong}~\bibnamefont{Xu}},
  \bibinfo{author}{\bibfnamefont{H.}~\bibnamefont{Hiraka}},
  \bibinfo{author}{\bibfnamefont{Z.}~\bibnamefont{Zhong}},
  \bibinfo{author}{\bibfnamefont{H.}~\bibnamefont{Luo}},
  \bibinfo{author}{\bibfnamefont{X.}~\bibnamefont{Zhao}},
  \bibinfo{author}{\bibfnamefont{D.}~\bibnamefont{Viehland}},
  \bibinfo{author}{\bibfnamefont{R.~J.} \bibnamefont{Birgeneau}},
  \bibnamefont{and}
  \bibinfo{author}{\bibfnamefont{G.} \bibnamefont{Shirane}},
  \bibinfo{journal}{Phys. Rev. B}
  \textbf{\bibinfo{volume}{73}}, \bibinfo{pages}{064107}
  (\bibinfo{year}{2006}).

\bibitem[{\citenamefont{{Guangyong Xu}
  et~al.}(2006{\natexlab{a}})\citenamefont{{Guangyong Xu}, Gehring, and
  Shirane}}]{xu:104110}
\bibinfo{author}{\bibnamefont{{Guangyong Xu}}},
  \bibinfo{author}{\bibfnamefont{P.~M.} \bibnamefont{Gehring}},
  \bibnamefont{and} \bibinfo{author}{\bibfnamefont{G.}~\bibnamefont{Shirane}},
  \bibinfo{journal}{Phys. Rev. B} \textbf{\bibinfo{volume}{74}},
  \bibinfo{pages}{104110} (\bibinfo{year}{2006}{\natexlab{a}}).

\bibitem[{\citenamefont{{Guangyong Xu}
  et~al.}(2006{\natexlab{b}})\citenamefont{{Guangyong Xu}, Zhong, Bing, Ye, and
  Shirane}}]{Xu_nm1}
\bibinfo{author}{\bibnamefont{{Guangyong Xu}}},
  \bibinfo{author}{\bibfnamefont{Z.}~\bibnamefont{Zhong}},
  \bibinfo{author}{\bibfnamefont{Y.}~\bibnamefont{Bing}},
  \bibinfo{author}{\bibfnamefont{Z.-G.} \bibnamefont{Ye}}, \bibnamefont{and}
  \bibinfo{author}{\bibfnamefont{G.}~\bibnamefont{Shirane}},
  \bibinfo{journal}{Nature Mater.} \textbf{\bibinfo{volume}{5}},
  \bibinfo{pages}{134} (\bibinfo{year}{2006}{\natexlab{b}}).

\bibitem[{\citenamefont{Gehring et~al.}(2004)\citenamefont{Gehring, Ohwada, and
  Shirane}}]{PZN-8PT}
\bibinfo{author}{\bibfnamefont{P.~M.} \bibnamefont{Gehring}},
  \bibinfo{author}{\bibfnamefont{K.}~\bibnamefont{Ohwada}}, \bibnamefont{and}
  \bibinfo{author}{\bibfnamefont{G.}~\bibnamefont{Shirane}},
  \bibinfo{journal}{Phys. Rev. B} \textbf{\bibinfo{volume}{70}},
  \bibinfo{pages}{014110} (\bibinfo{year}{2004}).

\bibitem[{\citenamefont{Jaffe et~al.}(1971)\citenamefont{Jaffe, Cook, and
  Jaffe}}]{ferrobook}
\bibinfo{author}{\bibfnamefont{B.}~\bibnamefont{Jaffe}},
  \bibinfo{author}{\bibfnamefont{W.~R.} \bibnamefont{Cook}}, \bibnamefont{and}
  \bibinfo{author}{\bibfnamefont{H.}~\bibnamefont{Jaffe}},
  \emph{\bibinfo{title}{Piezoelectric ceramics}} (\bibinfo{publisher}{Academic
  Press, London and New York}, \bibinfo{year}{1971}).

\bibitem[{\citenamefont{Kuwata et~al.}(1981)\citenamefont{Kuwata, Uchino, and
  Nomura}}]{PZN_phase2}
\bibinfo{author}{\bibfnamefont{J.}~\bibnamefont{Kuwata}},
  \bibinfo{author}{\bibfnamefont{K.}~\bibnamefont{Uchino}}, \bibnamefont{and}
  \bibinfo{author}{\bibfnamefont{S.}~\bibnamefont{Nomura}},
  \bibinfo{journal}{Ferroelectrics} \textbf{\bibinfo{volume}{37}},
  \bibinfo{pages}{579} (\bibinfo{year}{1981}).

\bibitem[{\citenamefont{Shrout et~al.}(1990)\citenamefont{Shrout, Chang, Kim,
  and Markgraf}}]{shrout1}
\bibinfo{author}{\bibfnamefont{T.}~\bibnamefont{Shrout}},
  \bibinfo{author}{\bibfnamefont{N.~P.} \bibnamefont{Chang}},
  \bibinfo{author}{\bibfnamefont{N.}~\bibnamefont{Kim}}, \bibnamefont{and}
  \bibinfo{author}{\bibfnamefont{S.}~\bibnamefont{Markgraf}},
  \bibinfo{journal}{Ferroelectr. Lett. Sect.} \textbf{\bibinfo{volume}{12}},
  \bibinfo{pages}{63} (\bibinfo{year}{1990}).

\bibitem[{\citenamefont{{Guangyong Xu, Z. Zhong}
  et~al.}(2004)\citenamefont{{Guangyong Xu, Z. Zhong}, Hiraka, and
  Shirane}}]{GXU3D}
\bibinfo{author}{\bibnamefont{{Guangyong Xu, Z. Zhong}}},
  \bibinfo{author}{\bibfnamefont{H.}~\bibnamefont{Hiraka}}, \bibnamefont{and}
  \bibinfo{author}{\bibfnamefont{G.}~\bibnamefont{Shirane}},
  \bibinfo{journal}{Phys. Rev. B} \textbf{\bibinfo{volume}{70}},
  \bibinfo{pages}{174109} (\bibinfo{year}{2004}).

\bibitem[{\citenamefont{Hirota et~al.}(2002)\citenamefont{Hirota, Ye, Wakimoto,
  Gehring, and Shirane}}]{PMN_diffuse}
\bibinfo{author}{\bibfnamefont{K.}~\bibnamefont{Hirota}},
  \bibinfo{author}{\bibfnamefont{Z.-G.} \bibnamefont{Ye}},
  \bibinfo{author}{\bibfnamefont{S.}~\bibnamefont{Wakimoto}},
  \bibinfo{author}{\bibfnamefont{P.~M.} \bibnamefont{Gehring}},
  \bibnamefont{and} \bibinfo{author}{\bibfnamefont{G.}~\bibnamefont{Shirane}},
  \bibinfo{journal}{Phys. Rev. B} \textbf{\bibinfo{volume}{65}},
  \bibinfo{pages}{104105} (\bibinfo{year}{2002}).

\bibitem[{\citenamefont{Noheda et~al.}(2002)\citenamefont{Noheda, Cox, Shirane,
  Gao, and Ye}}]{PMN_phase}
\bibinfo{author}{\bibfnamefont{B.}~\bibnamefont{Noheda}},
  \bibinfo{author}{\bibfnamefont{D.~E.} \bibnamefont{Cox}},
  \bibinfo{author}{\bibfnamefont{G.}~\bibnamefont{Shirane}},
  \bibinfo{author}{\bibfnamefont{J.}~\bibnamefont{Gao}}, \bibnamefont{and}
  \bibinfo{author}{\bibfnamefont{Z.-G.} \bibnamefont{Ye}},
  \bibinfo{journal}{Phys. Rev. B} \textbf{\bibinfo{volume}{66}},
  \bibinfo{pages}{054104} (\bibinfo{year}{2002}).

\bibitem[{\citenamefont{{Hu Cao} et~al.}(2005)\citenamefont{{Hu Cao}, {Feiming
  Bai}, {Jiefang Li}, Viehland, {Guangyong Xu}, Hiraka, and Shirane}}]{hucao1}
\bibinfo{author}{\bibnamefont{{Hu Cao}}},
  \bibinfo{author}{\bibnamefont{{Feiming Bai}}},
  \bibinfo{author}{\bibnamefont{{Jiefang Li}}},
  \bibinfo{author}{\bibfnamefont{D.}~\bibnamefont{Viehland}},
  \bibinfo{author}{\bibnamefont{{Guangyong Xu}}},
  \bibinfo{author}{\bibfnamefont{H.}~\bibnamefont{Hiraka}}, \bibnamefont{and}
  \bibinfo{author}{\bibfnamefont{G.}~\bibnamefont{Shirane}},
  \bibinfo{journal}{J. Appl. Phys} \textbf{\bibinfo{volume}{97}},
  \bibinfo{pages}{094101} (\bibinfo{year}{2005}).

\bibitem[{\citenamefont{Gvasaliya et~al.}(2007)\citenamefont{Gvasaliya,
  Roessli, Cowley, Kojima, and Lushnikov}}]{gvasaliya}
\bibinfo{author}{\bibfnamefont{S.~N.} \bibnamefont{Gvasaliya}},
  \bibinfo{author}{\bibfnamefont{B.}~\bibnamefont{Roessli}},
  \bibinfo{author}{\bibfnamefont{R.~A.} \bibnamefont{Cowley}},
  \bibinfo{author}{\bibfnamefont{S.}~\bibnamefont{Kojima}}, \bibnamefont{and}
  \bibinfo{author}{\bibfnamefont{S.~G.} \bibnamefont{Lushnikov}},
  \bibinfo{journal}{J. Phys.:Condens. Matter} \textbf{\bibinfo{volume}{19}},
  \bibinfo{pages}{016219} (\bibinfo{year}{2007}).

\bibitem[{\citenamefont{{Guangyong Xu} et~al.}(2005)\citenamefont{{Guangyong
  Xu}, Gehring, and Shirane}}]{Xu_new}
\bibinfo{author}{\bibnamefont{{Guangyong Xu}}},
  \bibinfo{author}{\bibfnamefont{P.~M.} \bibnamefont{Gehring}},
  \bibnamefont{and} \bibinfo{author}{\bibfnamefont{G.}~\bibnamefont{Shirane}},
  \bibinfo{journal}{Phys. Rev. B} \textbf{\bibinfo{volume}{72}},
  \bibinfo{pages}{214106} (\bibinfo{year}{2005}).

\bibitem[{\citenamefont{Noheda et~al.}(2001)\citenamefont{Noheda, Cox, Shirane,
  Park, Cross, and Zhong}}]{skin1}
\bibinfo{author}{\bibfnamefont{B.}~\bibnamefont{Noheda}},
  \bibinfo{author}{\bibfnamefont{D.~E.} \bibnamefont{Cox}},
  \bibinfo{author}{\bibfnamefont{G.}~\bibnamefont{Shirane}},
  \bibinfo{author}{\bibfnamefont{S.-E.} \bibnamefont{Park}},
  \bibinfo{author}{\bibfnamefont{L.~E.} \bibnamefont{Cross}}, \bibnamefont{and}
  \bibinfo{author}{\bibfnamefont{Z.}~\bibnamefont{Zhong}},
  \bibinfo{journal}{Phys. Rev. Lett.} \textbf{\bibinfo{volume}{86}},
  \bibinfo{pages}{3891} (\bibinfo{year}{2001}).

\bibitem[{\citenamefont{Vanderbilt and Cohen}(2001)}]{vanderbilt}
\bibinfo{author}{\bibfnamefont{D.}~\bibnamefont{Vanderbilt}} \bibnamefont{and}
  \bibinfo{author}{\bibfnamefont{M.~H.} \bibnamefont{Cohen}},
  \bibinfo{journal}{Phys. Rev. B} \textbf{\bibinfo{volume}{63}},
  \bibinfo{pages}{094108} (\bibinfo{year}{2001}).

\end{thebibliography}

\end{document}